# Toward Practical $N^2$ Monte Carlo: the Marginal Particle Filter

**Mike Klaas**
Computer Science Dept.
University of British Columbia
klaas@cs.ubc.ca

**Nando de Freitas**
Computer Science Dept.
University of British Columbia
nando@cs.ubc.ca

**Arnaud Doucet**
Computer Science and Stat. Depts.
University of British Columbia
arnaud@cs.ubc.ca

## Abstract

Sequential Monte Carlo techniques are useful for state estimation in non-linear, non-Gaussian dynamic models. These methods allow us to approximate the joint posterior distribution using sequential importance sampling. In this framework, the dimension of the target distribution grows with each time step, thus it is necessary to introduce some resampling steps to ensure that the estimates provided by the algorithm have a reasonable variance. In many applications, we are only interested in the marginal filtering distribution which is defined on a space of fixed dimension. We present a Sequential Monte Carlo algorithm called the Marginal Particle Filter which operates directly on the marginal distribution, hence avoiding having to perform importance sampling on a space of growing dimension. Using this idea, we also derive an improved version of the auxiliary particle filter. We show theoretic and empirical results which demonstrate a reduction in variance over conventional particle filtering, and present techniques for reducing the cost of the marginal particle filter with $N$ particles from $O(N^2)$ to $O(N \log N)$.

## 1 Introduction

Bayesian state estimation is ubiquitous in the AI community, being one of the most popular techniques for performing inference in dynamic models. Examples of its use include tracking, diagnosis, and control. An unobserved signal (latent states $\mathbf{x}_t \in \mathcal{X}$) is assumed to exist, and evolves according to a (typically Markovian) dynamic model. Additionally, Bayesian filtering assumes the existence of observations ($\mathbf{y}_t$) which are conditionally independent given the process. An observation model specifies the generation of observations given a specified latent state. Of interest is estimating a distribution of the latent state at time $t$ given the observations up to that time; this is known as the filtering distribution $p(\mathbf{x}_t|\mathbf{y}_{1:t})$.

In some cases this model can be solved using exact inference, for instance using the Kalman or HMM filters. Unfortunately, real-world models are rarely simple enough to be solved exactly, often containing non-linearity, non-Gaussianity, or hybrid combinations of discrete and continuous variables which lead to high-dimensional intractable integrals. Since these integrals cannot be solved analytically, approximation techniques are required.

One of the most successful and popular approximation techniques is Sequential Monte Carlo (SMC), which is referred to as Particle Filtering (PF) in the Bayesian filtering domain [2, 13, 3]. In its most basic form, particle filters work by starting with a sample from the posterior at time $t-1$, *predicting* the state at time $t$, then *updating* the importance weights based on the observation $\mathbf{y}_t$. These samples form an approximation of the joint density $p(\mathbf{x}_{1:t}|\mathbf{y}_{1:t})$ at time $t$. Often, however, it is the *filtering* distribution $p(\mathbf{x}_t|\mathbf{y}_{1:t})$ that is desired. This is approximated by dropping samples of the states $\mathbf{x}_1 \ldots \mathbf{x}_{t-1}$ at time $t$ (often implicitly). The tradeoff for doing the importance sampling sequentially when the marginal is of interest is that the particle filter is performing importance sampling in a higher-dimensional state space $\mathcal{X}^t$ than is necessary, which results in higher variance of the importance weights and requires the use of resampling steps.

In this paper, we develop two particle filtering algorithms which performs importance sampling directly in the *marginal space* $\mathcal{X}$ of $p(\mathbf{x}_t|\mathbf{y}_{1:t})$. We show using synthetic and real-world experiments that these algorithms improve significantly over Sequential Importance Sampling/Resampling (SIR) and Auxiliary Particle Filtering (ASIR) in terms of importance weight variance. The main disadvantage of doing importance sampling in the marginal space is its $O(N^2)$ cost, where $N$ is the number of particles used. We demonstrate how this cost can be reduced to $O(N \log N)$ or even $O(N)$ using methods from $N$-body learning [4, 9].

## 2 Bayesian filtering with SMC

The unobserved signal $\{\mathbf{x}_t\}$, $\mathbf{x}_t \in \mathcal{X}$, is modelled as a *Markov process* of initial distribution $p(\mathbf{x}_1)$ and transition prior $p(\mathbf{x}_t|\mathbf{x}_{t-1})$. The observations $\{\mathbf{y}_t\}$, $\mathbf{y}_t \in \mathcal{Y}$, are assumed to be conditionally independent given the process $\{\mathbf{x}_t\}$ and of marginal distribution $p(\mathbf{y}_t|\mathbf{x}_t)$. Hence, the model is described by

$$p(\mathbf{x}_t|\mathbf{x}_{t-1}) \quad t \geq 1, \text{ and}$$
$$p(\mathbf{y}_t|\mathbf{x}_t) \quad t \geq 1.$$

We denote by $\mathbf{x}_{1:t} \triangleq \{\mathbf{x}_1, ..., \mathbf{x}_t\}$ and $\mathbf{y}_{1:t} \triangleq \{\mathbf{y}_1, ..., \mathbf{y}_t\}$, respectively, the signal and the observations up to time $t$, and define $p(\mathbf{x}_1|\mathbf{x}_0) \triangleq p(\mathbf{x}_1)$ for notational convenience.

Our aim is to estimate sequentially in time the *filtering distribution* $p(\mathbf{x}_t|\mathbf{y}_{1:t})$ and the expectations

$$I(f_t) = \mathbb{E}_{p(\mathbf{x}_t|\mathbf{y}_{1:t})}[f_t(\mathbf{x}_t)] \triangleq \int f_t(\mathbf{x}_t) p(\mathbf{x}_t|\mathbf{y}_{1:t}) d\mathbf{x}_t \quad (1)$$

for some function of interest $f_t : \mathcal{X} \to \mathbb{R}^{n_{f_t}}$ integrable with respect to $p(\mathbf{x}_t|\mathbf{y}_{1:t})$. Examples of appropriate functions include the conditional mean, in which case $f_t(\mathbf{x}_t) = \mathbf{x}_t$, or the conditional covariance of $\mathbf{x}_t$ where $f_t(\mathbf{x}_t) = \mathbf{x}_t\mathbf{x}_t^{\mathsf{T}} - \mathbb{E}_{p(\mathbf{x}_t|\mathbf{y}_{1:t})}[\mathbf{x}_t]\mathbb{E}_{p(\mathbf{x}_t|\mathbf{y}_{1:t})}^{\mathsf{T}}[\mathbf{x}_t]$.

### 2.1 Sequential importance sampling

If we had a set of samples (or *particles*) $\{\mathbf{x}_t^{(i)}\}_{i=1}^N$ from $p(\mathbf{x}_t|\mathbf{y}_{1:t})$, we could approximate the distribution with the Monte Carlo estimator

$$\widehat{p}(d\mathbf{x}_t|\mathbf{y}_{1:t}) = \frac{1}{N}\sum_{i=1}^N \delta_{\mathbf{x}_t^{(i)}}(d\mathbf{x}_t)$$

where $\delta_{\mathbf{x}_t^{(i)}}(d\mathbf{x}_t)$ denotes the delta Dirac function. This can be used to approximate the expectations of interest in equation (1) with

$$\widehat{I}(f_t) = \frac{1}{N}\sum_{i=1}^N f_t\left(\mathbf{x}_t^{(i)}\right).$$

This estimate converges almost surely to the true expectation as $N$ goes to infinity. Unfortunately, one cannot easily sample from the marginal distribution $p(\mathbf{x}_t|\mathbf{y}_{1:t})$ directly. Instead, we draw particles from $p(\mathbf{x}_{1:t}|\mathbf{y}_{1:t})$ and samples $\mathbf{x}_{1:t-1}$ are ignored. To draw samples from $p(\mathbf{x}_{1:t}|\mathbf{y}_{1:t})$, we sample from a proposal distribution $q$ and weight the particles according to the following importance ratio:

$$w_t(\mathbf{x}_{1:t}) = \frac{p(\mathbf{x}_{1:t}|\mathbf{y}_{1:t})}{q(\mathbf{x}_{1:t}|\mathbf{y}_{1:t})}$$

The proposal distribution is constructed sequentially

$$q(\mathbf{x}_{1:t}|\mathbf{y}_{1:t}) = q(\mathbf{x}_{1:t-1}|\mathbf{y}_{1:t-1})q(\mathbf{x}_t|\mathbf{y}_t, \mathbf{x}_{t-1})$$

and, hence, the importance weights can be updated recursively

$$w_t(\mathbf{x}_{1:t}) = \frac{p(\mathbf{x}_{1:t}|\mathbf{y}_{1:t})}{p(\mathbf{x}_{1:t-1}|\mathbf{y}_{1:t-1})\,q(\mathbf{x}_t|\mathbf{y}_t,\mathbf{x}_{t-1})}w_{t-1}(\mathbf{x}_{1:t-1}). \quad (2)$$

Given a set of $N$ particles $\{\mathbf{x}_{1:t-1}^{(i)}\}_{i=1}^N$, we obtain a set of particles $\{\mathbf{x}_{1:t}^{(i)}\}_{i=1}^N$ by sampling from $q(\mathbf{x}_t|\mathbf{y}_t, \mathbf{x}_{t-1}^{(i)})$ and applying the weights of equation (2).

The familiar particle filtering equations for this model are obtained by remarking that

$$p(\mathbf{x}_{1:t}|\mathbf{y}_{1:t}) \propto p(\mathbf{x}_{1:t}, \mathbf{y}_{1:t}) = \prod_{k=1}^t p(\mathbf{y}_k|\mathbf{x}_k)p(\mathbf{x}_k|\mathbf{x}_{k-1}),$$

given which, equation (2) becomes

$$\widetilde{w}_t^{(i)} \propto \frac{p(\mathbf{y}_t|\mathbf{x}_t^{(i)})p(\mathbf{x}_t^{(i)}|\mathbf{x}_{t-1}^{(i)})}{q\left(\mathbf{x}_t^{(i)}\middle|\mathbf{y}_t,\mathbf{x}_{t-1}^{(i)}\right)}\widetilde{w}_{t-1}^{(i)}.$$

This iterative scheme produces a weighted measure $\{\mathbf{x}_{1:t}^{(i)}, w_t^{(i)}\}_{i=1}^N$, where $w_t^{(i)} = \widetilde{w}_t^{(i)}/\sum_j \widetilde{w}_t^{(j)}$, and is known as Sequential Importance Sampling (SIS). A resampling (selection) step may be included at this point that chooses the fittest particles (this is the SIR algorithm [8]).

Figure 2 contains pseudo-code for the algorithm. Note this is the procedure in common use by practitioners. It can be deceptive: although only the state $\mathbf{x}_t$ is being updated every round, the algorithm (as we have presently derived) is nonetheless importance sampling in the *joint* space $\mathcal{X}^t$.

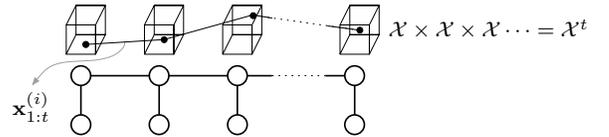

Figure 1: Sequential importance sampling.

## 3 Marginal Particle Filter

Sequential importance sampling estimates $p(\mathbf{x}_{1:t}|\mathbf{y}_{1:t})$ by taking an estimate of $p(\mathbf{x}_{1:t-1}|\mathbf{y}_{1:t-1})$ and augmenting it with a new sample $\mathbf{x}_t$ at time $t$. Each particle at time $t$ is a draw over the joint space $p(\mathbf{x}_{1:t}|\mathbf{y}_{1:t})$, sampled sequentially, thus can be thought of as a *path* through the state space at times $1 \ldots t$ (Figure 1). At each time step, the dimension of the sampled paths is increased by the dimension of the state space at time $t$, quickly resulting in a very high dimensional space. The sequential nature of the algorithm means that the variance is high, leading to most paths having vanishingly small probability. This problem is known as *degeneracy* of the weights, and usually leads to weights whose variance tends to increase without bound.

*Sequential importance sampling step*

- For $i = 1, ..., N$, sample from the proposal

$$\mathbf{x}_t^{(i)} \sim q\left(\mathbf{x}_t \big| \mathbf{y}_t, \mathbf{x}_{t-1}^{(i)}\right)$$

- For $i = 1, ..., N$, evaluate the importance weights

$$\widetilde{w}_t^{(i)} = \frac{p\left(\mathbf{y}_t \big| \mathbf{x}_t^{(i)}\right) p\left(\mathbf{x}_t^{(i)} \big| \mathbf{x}_{t-1}^{(i)}\right)}{q\left(\mathbf{x}_t^{(i)} \big| \mathbf{y}_t, \mathbf{x}_{t-1}^{(i)}\right)} \widetilde{w}_{t-1}^{(i)}$$

- Normalise the importance weights

$$w_t^{(i)} = \frac{\widetilde{w}_t^{(i)}}{\sum_j^N \widetilde{w}_t^{(j)}}$$

*Selection step*

- Resample the discrete weighted measure $\left\{\mathbf{x}_t^{(i)}, w_t^{(i)}\right\}_{i=1}^N$ to obtain an unweighted measure $\left\{\mathbf{x}_t^{(i)}, \frac{1}{N}\right\}_{i=1}^N$ of $N$ new particles.

Figure 2: Particle filtering algorithm at time $t$.

One strategy employed is to use a *resampling* step after updating the particle weights to multiply particles (paths) with high weight and prune particles with negligible weight (the SIR algorithm).

The Marginal Particle Filter (MPF) uses a somewhat more principled approach. We perform particle filtering *directly* on the marginal distribution $p(\mathbf{x}_t|\mathbf{y}_{1:t})$ instead of on the joint space.

The predictive density is obtained by marginalizing

$$p(\mathbf{x}_t|\mathbf{y}_{1:t-1}) = \int p(\mathbf{x}_t|\mathbf{x}_{t-1}) p(\mathbf{x}_{t-1}|\mathbf{y}_{1:t-1}) d\mathbf{x}_{t-1} \quad (3)$$

hence, the filtering update becomes

$$p(\mathbf{x}_t|\mathbf{y}_{1:t}) \propto p(\mathbf{y}_t|\mathbf{x}_t) p(\mathbf{x}_t|\mathbf{y}_{1:t-1})$$
$$= p(\mathbf{y}_t|\mathbf{x}_t) \int p(\mathbf{x}_t|\mathbf{x}_{t-1}) p(\mathbf{x}_{t-1}|\mathbf{y}_{1:t-1}) d\mathbf{x}_{t-1}.$$

The integral in equation (3) is generally not solvable analytically, but since we have a particle approximation of $p(\mathbf{x}_{t-1}|\mathbf{y}_{1:t-1})$ $\left(\text{namely, } \left\{\mathbf{x}_{t-1}^{(i)}, w_{t-1}^{(i)}\right\}\right)$, we can approximate (3) as the weighted kernel estimate $\sum_{j=1}^N w_{t-1}^{(j)} p\left(\mathbf{x}_t \big| \mathbf{x}_{t-1}^{(j)}\right)$.

While we are free to choose any proposal distribution that has appropriate support, it is convenient to assume that the proposal takes a similar form, namely

$$q(\mathbf{x}_t|\mathbf{y}_{1:t}) = \sum_{j=1}^N w_{t-1}^{(j)} q\left(\mathbf{x}_t \big| \mathbf{y}_t, \mathbf{x}_{t-1}^{(j)}\right). \quad (4)$$

The importance weights are now on the marginal space

$$w_t = \frac{p(\mathbf{x}_t|\mathbf{y}_{1:t})}{q(\mathbf{x}_t|\mathbf{y}_{1:t})}.$$

Pseudo-code for the algorithm is given in Figure 3.

*Marginal Particle Filter (MPF)*

- For $i = 1, ..., N$, sample from the proposal using stratified sampling

$$\mathbf{x}_t^{(i)} \sim \sum_{j=1}^N w_{t-1}^{(j)} q\left(\mathbf{x}_t \big| \mathbf{y}_t, \mathbf{x}_{t-1}^{(j)}\right)$$

- For $i = 1, ..., N$, evaluate the importance weights

$$\widetilde{w}_t^{(i)} = \frac{p\left(\mathbf{y}_t \big| \mathbf{x}_t^{(i)}\right) \sum_{j=1}^N w_{t-1}^{(j)} p\left(\mathbf{x}_t^{(i)} \big| \mathbf{x}_{1:t-1}^{(j)}\right)}{\sum_{j=1}^N w_{t-1}^{(j)} q\left(\mathbf{x}_t^{(i)} \big| \mathbf{y}_t, \mathbf{x}_{t-1}^{(j)}\right)}$$

- Normalize the importance weights

$$w_t^{(i)} = \frac{\widetilde{w}_t^{(i)}}{\sum_j \widetilde{w}_t^{(j)}}$$

Figure 3: The MPF algorithm at time $t$.

### 3.1 Auxiliary Variable MPF

The auxiliary particle filter (ASIR), introduced by Pitt and Shepard [11, 1], is designed to improve the performance of sequential Monte Carlo in models with peaked likelihoods (which is another source of importance weight variance). In this section, we derive an algorithm that combines both approaches.

When the likelihood is narrow, it is desirable to choose a proposal distribution that samples particles which will be in high-probability regions of the observation model. The auxiliary particle filter works by re-weighting the particles at time $t-1$ to boost them in these regions.

We are interested in sampling from the following target distribution

$$\widehat{p}(\mathbf{x}_t|\mathbf{y}_{1:t}) \propto p(\mathbf{y}_t|\mathbf{x}_t) \sum_{j=1}^N w_{t-1}^{(j)} p\left(\mathbf{x}_t \big| \mathbf{x}_{t-1}^{(j)}\right) \quad (5)$$
$$= \sum_{j=1}^N w_{t-1}^{(j)} p\left(\mathbf{y}_t \big| \mathbf{x}_{t-1}^{(j)}\right) p\left(\mathbf{x}_t \big| \mathbf{x}_{t-1}^{(j)}, \mathbf{y}_t\right).$$

The auxiliary PF uses the following joint distribution:

$$\widehat{p}(k, \mathbf{x}_t | \mathbf{y}_{1:t}) \propto w_{t-1}^{(k)} p\left(\mathbf{y}_t \Big| \mathbf{x}_{t-1}^{(k)}\right) p\left(\mathbf{x}_t \Big| \mathbf{x}_{t-1}^{(k)}, \mathbf{y}_t\right).$$

$k$ is known as an *auxiliary variable* and is an index into the mixture of equation (5). Thus,

$$\widehat{p}(k | \mathbf{y}_{1:t}) \propto w_{t-1}^{(k)} p\left(\mathbf{y}_t \Big| \mathbf{x}_{t-1}^{(k)}\right). \quad (6)$$

$$= w_{t-1}^{(k)} \int p(\mathbf{y}_t | \mathbf{x}_t) p\left(\mathbf{x}_t \Big| \mathbf{x}_{t-1}^{(k)}\right) d\mathbf{x}_t$$

Since the exact evaluation of (6) is usually impossible, we approximate this via what is known as a *simulation* step. For each index $k$ at time $t-1$, we choose $\mu_t^{(k)}$ associated with the distribution $p(\mathbf{x}_t | \mathbf{x}_{t-1}^{(k)})$ in some deterministic fashion ($\mu_t^{(k)}$ could be the expected value, for instance). We define the *simulation weight*[1] for index $k$ to be

$$\lambda_{t-1}^{(k)} \triangleq \frac{w_{t-1}^{(k)} p\left(\mathbf{y}_t \Big| \mu_t^{(k)}\right)}{\sum_{j=1}^N w_{t-1}^{(j)} p\left(\mathbf{y}_t \Big| \mu_t^{(j)}\right)}.$$

Using these weights, the auxiliary particle filter defines the following joint proposal distribution:

$$q(k, \mathbf{x}_t | \mathbf{y}_{1:t}) = q(k | \mathbf{y}_{1:t}) q(\mathbf{x}_t | \mathbf{y}_{1:t}, k)$$

where

$$q(k | \mathbf{y}_{1:t}) = \lambda_{t-1}^{(k)},$$

$$q(\mathbf{x}_t | \mathbf{y}_{1:t}, k) = q\left(\mathbf{x}_t \Big| \mathbf{x}_{t-1}^{(k)}, \mathbf{y}_t\right).$$

The importance weight is given by

$$w(k, \mathbf{x}_t) = \frac{\widehat{p}(k, \mathbf{x}_t | \mathbf{y}_{1:t})}{q(k, \mathbf{x}_t | \mathbf{y}_{1:t})} \quad (7)$$

$$\propto \frac{w_{t-1}^{(k)} p\left(\mathbf{y}_t \Big| \mathbf{x}_{t-1}^{(k)}\right) p\left(\mathbf{x}_t \Big| \mathbf{x}_{t-1}^{(k)}, \mathbf{y}_t\right)}{\lambda_{t-1}^{(k)} q\left(\mathbf{x}_t \Big| \mathbf{x}_{t-1}^{(k)}, \mathbf{y}_t\right)}.$$

In the marginal particle filter, we use the same importance distribution but instead of performing importance sampling between $\widehat{p}(k, \mathbf{x}_t | \mathbf{y}_{1:t})$ and $q(k, \mathbf{x}_t | \mathbf{y}_{1:t})$, we *directly* perform importance sampling between $\widehat{p}(\mathbf{x}_t | \mathbf{y}_{1:t})$ and $q(\mathbf{x}_t | \mathbf{y}_{1:t})$ to compute the weights

$$w(\mathbf{x}_t) = \frac{\widehat{p}(\mathbf{x}_t | \mathbf{y}_{1:t})}{q(\mathbf{x}_t | \mathbf{y}_{1:t})} \quad (8)$$

$$\propto \frac{\sum_{j=1}^N w_{t-1}^{(j)} p\left(\mathbf{y}_t \Big| \mathbf{x}_{t-1}^{(j)}\right) p\left(\mathbf{x}_t \Big| \mathbf{x}_{t-1}^{(j)}, \mathbf{y}_t\right)}{\sum_{j=1}^N \lambda_{t-1}^{(j)} q\left(\mathbf{x}_t \Big| \mathbf{x}_{t-1}^{(j)}, \mathbf{y}_t\right)}.$$

This leads to the auxiliary marginal particle filter (AMPF) which is described in Figure 4.

---

[1] To prevent this step from introducing bias, the simulation weights are chosen independently from $\mathbf{x}_t^{(i)}$.

---

*Auxiliary Marginal Particle Filter (AMPF)*

- For $i = 1, \ldots, N$, choose simulation $\mu_t^{(i)}$ and calculate mixture weights

$$\mu_t^{(i)} \leftarrow_d p\left(\mathbf{x}_t \Big| \mathbf{x}_{t-1}^{(i)}\right)$$

$$\widetilde{\lambda}_{t-1}^{(i)} = w_{t-1}^{(i)} p\left(\mathbf{y}_t \Big| \mu_t^{(i)}\right) \quad \lambda_{t-1}^{(i)} = \frac{\widetilde{\lambda}_{t-1}^{(i)}}{\sum_j^N \widetilde{\lambda}_{t-1}^{(j)}}$$

- For $i = 1, \ldots, N$, sample from the proposal

$$\mathbf{x}_t^{(i)} \sim \sum_{j=1}^N \lambda_{t-1}^{(j)} q\left(\mathbf{x}_t \Big| \mathbf{y}_t, \mathbf{x}_{t-1}^{(j)}\right)$$

- For $i = 1, \ldots, N$, evaluate the importance weights

$$\widetilde{w}_t^{(i)} = \frac{p\left(\mathbf{y}_t \Big| \mathbf{x}_t^{(i)}\right) \sum_{j=1}^N w_{t-1}^{(j)} p\left(\mathbf{x}_t^{(i)} \Big| \mathbf{x}_{t-1}^{(j)}\right)}{\sum_{j=1}^N \lambda_{t-1}^{(j)} q\left(\mathbf{x}_t^{(i)} \Big| \mathbf{y}_t, \mathbf{x}_{t-1}^{(j)}\right)}$$

- Normalise the importance weights

$$w_t^{(i)} = \frac{\widetilde{w}_t^{(i)}}{\sum_{j=1}^N \widetilde{w}_t^{(j)}}$$

Figure 4: The AMPF algorithm at time $t$. The $\leftarrow_d$ symbol denotes the deterministic selection of a likely value from the distribution, such as the mean or a mode of the density.

We expect that performing importance sampling directly between the distributions of interest will lead to a reduction in variance. It is not hard to show that it can be no worse.

**Proposition 1.** *The variance of the AMPF importance sampling weights $w(\mathbf{x}_t)$ is less than or equal to ASIR's importance weights $w(k, \mathbf{x}_t)$.*

*Proof.* By the variance decomposition lemma, we have

$$var[w(k, \mathbf{x}_t)] = var[\mathbb{E}(w(k, \mathbf{x}_t) | \mathbf{x}_t)]$$
$$+ \mathbb{E}[var(w(k, \mathbf{x}_t) | \mathbf{x}_t)]$$
$$= var[w(\mathbf{x}_t)] + \mathbb{E}[var(w(k, \mathbf{x}_t) | \mathbf{x}_t)].$$

Hence, as

$$\mathbb{E}[var(w(k, \mathbf{x}_t) | \mathbf{x}_t)] \geq 0$$

it follows that

$$var[w(\mathbf{x}_t)] \leq var[w(k, \mathbf{x}_t)]. \quad \square$$

### 3.2 Discussion

Since the particles at time $t$ are sampled from a much lower dimensional space in the marginal filter algorithms, we expect that the variance in the weights will be significantly less than that of (A)SIR for the same number of particles.

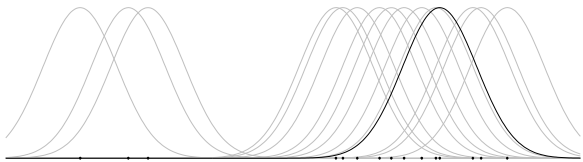

(a) (A)SIR samples a mixture component and uses this to compute importance sampling weights.

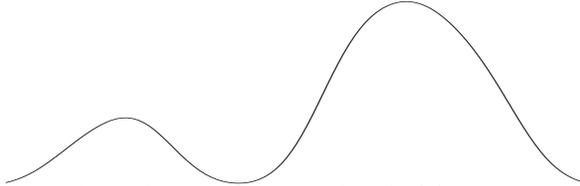

(b) Marginal filtering uses the entire mixture.

Figure 5: Predictive density $p(\mathbf{x}_t|\mathbf{y}_{1:t-1})$ in (A)SIR and Marginal PF. By using a single mixture component, the (A)SIR estimate ignores important details of the distribution; a particle lying in the left mode should be given less weight than one lying in the right mode.

Proposition 1 proves that it is no greater in the auxiliary variable setting, and a similar result holds for SIR.

Figure 5 demonstrates this with an example. Consider a multi-modal state estimate at time $t-1$, and a Gaussian transition prior. In (A)SIR, a particle's transition likelihood is relative to the tail end of a path (5(a)), while the marginal PF calculates the true marginal transition density (5(b)).

The marginal PF improves over (A)SIR whenever a particle has high marginal probability but low joint probability along its path. This can occur due to heavy-tailed importance distributions or models with narrow or misspecified transition priors. On the other hand, the improvement of MPF over SIR will not be very pronounced if the observation model is peaked (i.e., if likelihood is highly concentrated), as this will influence the importance weights more than the effect of sampling in the joint space. In these cases, AMPF should be used. Figure 6 demonstrates the two types of variance reduction.

Finally, it should be noted that sequential Monte Carlo applies to domains outside of Bayesian filtering, and an analogous marginal SMC algorithm can be straightforwardly derived in a general SMC context.

Note that the evaluation of the proposal (equation (4)) must be performed for each sample, thus both MPF and AMPF incur an $N^2$ cost. As we later show, this can be improved substantially.

### 3.3 Cases of equivalence

There is one case where SIR and marginal PF are equivalent. When the transition prior is used as the proposal dis-

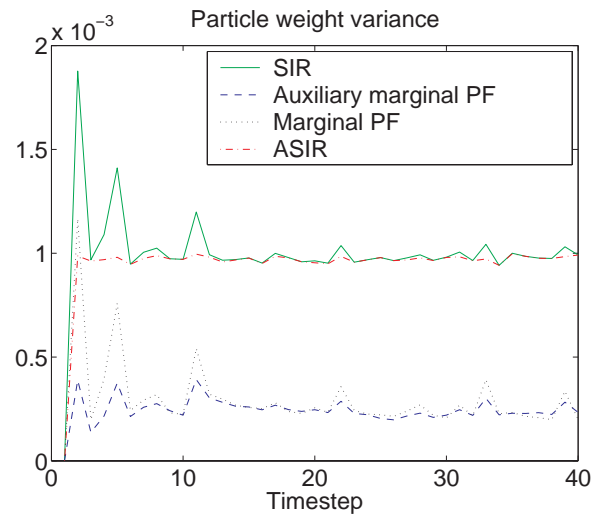

Figure 6: Importance weights variance reduction. "Spikes" in weight variance are caused by unlikely observations. ASIR (red) is successful in smoothing these occurrences when using SIR (green). The marginal PF (black) reduces overall variance by sampling in a smaller-dimensional space, but still suffers from spikes. AMPF (blue) gains the advantages of both approaches.

tribution, then the MPF weight update equation becomes:

$$w_t^{(i)} \propto \frac{p\left(\mathbf{y}_t\middle|\mathbf{x}_t^{(i)}\right) \sum_{j=1}^N w_{t-1}^{(j)} p\left(\mathbf{x}_t^{(i)}\middle|\mathbf{x}_{t-1}^{(j)}\right)}{\sum_{j=1}^N w_{t-1}^{(j)} p\left(\mathbf{x}_t^{(i)}\middle|\mathbf{x}_{t-1}^{(j)}\right)}$$
$$= p\left(\mathbf{y}_t\middle|\mathbf{x}_t^{(i)}\right).$$

When using SIR, particles are resampled after being weighted, but this is *precisely* equivalent to sampling from the marginal proposal distribution $\sum_j^N w_{t-1}^{(j)} p\left(\mathbf{x}_t\middle|\mathbf{x}_{t-1}^{(j)}\right)$. The SIR weight update equation is

$$w_t^{(i)} \propto \frac{p\left(\mathbf{y}_t\middle|\mathbf{x}_t^{(i)}\right) p\left(\mathbf{x}_t^{(i)}\middle|\mathbf{x}_{t-1}^{(i)}\right)}{p\left(\mathbf{x}_t^{(i)}\middle|\mathbf{x}_{t-1}^{(i)}\right)} w_{t-1}^{(i)} = p\left(\mathbf{y}_t\middle|\mathbf{x}_t^{(i)}\right),$$

since $w_{t-1}^{(i)}$ is set to $N^{-1}$ after resampling. In both cases, the conventional likelihood-weighted filter is recovered.

Similarly, when performing auxiliary filtering, it may be possible to sample exactly from the optimal proposal $q(\mathbf{x}_t|\mathbf{y}_t, k) = p(\mathbf{x}_t|\mathbf{y}_t, \mathbf{x}_t^{(k)})$ *and* exactly evaluate $p(\mathbf{y}_t|\mathbf{x}_{t-1}^{(k)})$ (equation (6)). In this case the importance weight variance is zero, thus the marginal particle filter cannot bring any improvement.

## 4 Efficient Implementation

The marginal particle filter as presented has a significant disadvantage: it has $O(N^2)$ complexity. In this section,

we show how to apply powerful techniques from $N$-body learning [4] to reduce this cost to $O(N \log N)$, and in some cases $O(N)$. $N$-body problems involve a set of *sources* $X \triangleq \{x_j\}$ with associated weights $\{\omega_j\}$, a set of *targets* $Y \triangleq \{y_i\}$,[2] and an *influence function* $K(x_j, y_i)$.

The goal is to find the *influence* $q_i$ at each target $y_i$, ie.

$$\forall i, \qquad q_i = \sum_{j=1}^{N} \omega_j K(x_j, y_i)$$

to within a specified error $\epsilon$. This is the weighted *Kernel Density Estimation* (KDE) problem, and costs $O(N^2)$ when implemented directly. The mapping to marginal filtering is direct. For instance, equation (4) can be formulated as a weighted KDE problem as follows:

$$\{x_j\} \triangleq \left\{\mathbf{x}_{t-1}^{(j)}\right\}, \quad \{y_i\} \triangleq \left\{\mathbf{x}_t^{(i)}\right\}, \quad \text{and}$$
$$K(x_j, y_i) \triangleq q(\mathbf{x}_t = y_i | \mathbf{x}_{t-1} = x_j, \mathbf{Y}_t).$$

We give a brief overview of the methods involved. For an empirical comparison of the methods, see [9].

### 4.1 Dual-tree methods

Dual-tree recursion requires a kernel parameterized by a distance function: $K(x, y) \triangleq K(\delta(x, y))$. This encompasses most continuous densities and some discrete distributions (in the latter case, it usually depends on the parameters of the distribution). This method first builds a spacial tree[3] on *both* the source and target points. We can bound the kernel response of a node of source points to a node of target points based on the node-node distances (Figure 7). The bounds can be tightened by recursing on both trees. Gray and Moore [5] give a thorough exposition.

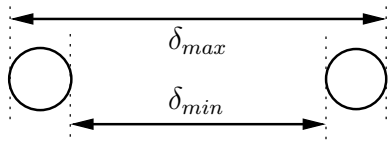

Figure 7: Dual-tree bounding.

### 4.2 Gaussian kernel methods

If the kernel is Gaussian, then the fast Gauss Transform [6] and Improved-FGT [14] can be used to perform weighted KDE in $O(N)$. These methods work by partitioning the space in which the data lie into clusters, and approximating the Gaussian sum from one cluster to another with a Hermite polynomial (FGT) or Taylor series expansion (IFGT). While the IFGT is generally faster, it is unknown how to tune the parameters to guarantee the error bound while maintaining acceptable performance.

---
[2]Note: $x_j$, $y_i$, and $\omega_j$ defined here are not to be confused with the $\mathbf{x}_t$, $\mathbf{y}_t$, and $w_t$ used earlier in the paper.

[3]A popular choice is a $kd$-tree, though metric trees are preferred in high dimensional spaces.

## 5 Experiments

We present several examples of using the Marginal Particle Filter for Bayesian filtering. We compare the algorithms in several respects: the error of the state estimate to the ground truth (when known); the variance of the importance weights; and *unique particle count*. High variance is indicative of degeneracy of the importance sampling weights, and affects the precision and variance of the estimator. Unique particle count is a measure of the diversity of the particles. The latter two are both important: it is trivial to construct an algorithm which performs well under either of these measures individually, but the construction will behave pathologically under the other measure.

### 5.1 Multi-modal non-linear time series

Consider the following reference model [2]:

$$x_t = \frac{x_{t-1}}{2} + \frac{25 x_{t-1}}{1 + x_{t-1}^2} + \cos(1.2t) + \mathcal{N}(0, \sigma_x)$$
$$y_t = \frac{x_t^2}{20} + \mathcal{N}(0, \sigma_y).$$

The posterior distribution is bi-modal and non-linear; this is a standard example of a difficult filtering workload.

| Method | RMS Error | variance | Weight var. |
|--------|-----------|----------|-------------|
| SIR    | 2.902     | 1.03     | 0.000163    |
| MPF    | 2.344     | 0.06     | 0.000025    |

Table 1: 1-D time series: RMS error and weight variance.

Table 1 and Figures 8, 9, and 10 summarize the results. The Marginal Particle Filter improves over SIR slightly in terms of RMSE, and produces a substantial reduction in importance weight variance.

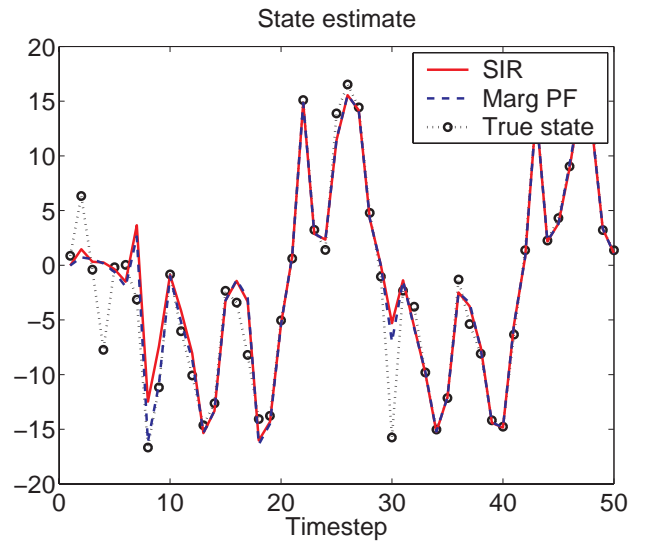

Figure 8: 1-D time series; state estimate.

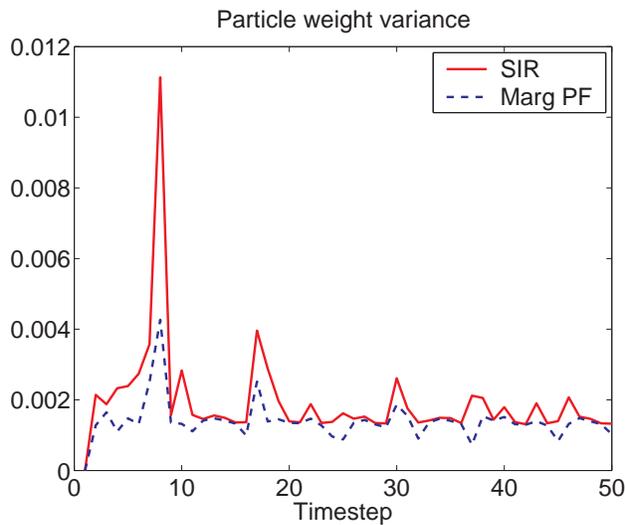

Figure 9: 1-D time series; variance of the importance weights. The spikes are the result of unlikely data—MPF does particularly better than SIR in these cases.

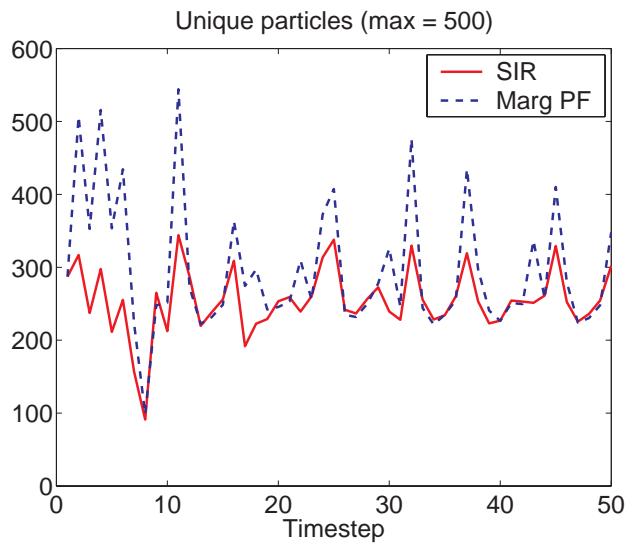

Figure 10: 1-D time series; unique particle count. Although the Marginal PF generally better diversity, an unlikely observation can still cause problems (such as at $t = 8$).

### 5.2 Stochastic Volatility

Monte Carlo methods are often applied to the analysis of the variance of financial returns as the models involved cannot be solved analytically. One commonly used model is stochastic volatility [7], which is summarized as:

$$y_t = \epsilon_t \beta \exp\{x_t/2\}$$
$$x_t = \phi x_{t-1} + \eta_t$$

where $\eta_t \sim \mathcal{N}(0, \sigma_\eta)$, $\epsilon_t \sim \mathcal{N}(0, 1)$, and $x_1 \sim \mathcal{N}(0, \sigma_\eta^2/(1-\phi^2))$. We analyze the weekday close of the U.K. Sterling/U.S. Dollar exchange rate from 1/10/81 to 28/6/85. There are 946 timesteps, but we limit analysis to the first 200 for readability, and use the model parameters fit to the data using MCMC in [7]. We use as proposal the transition prior with heavier tails to test the marginal filter's ability to compensate for poor proposals.

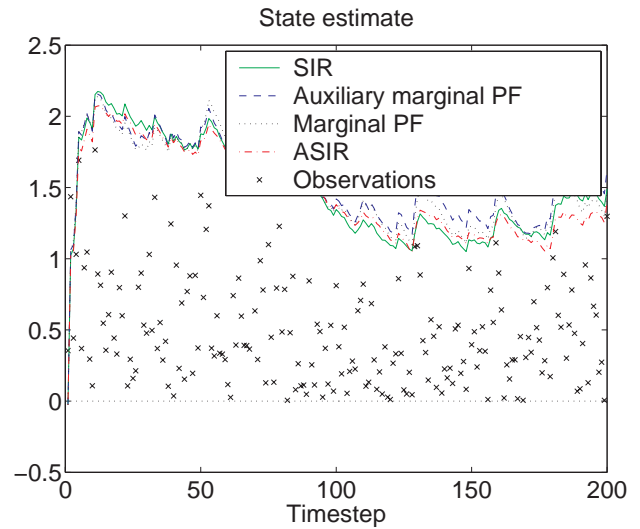

Figure 11: Stochastic volatility model; state estimate.

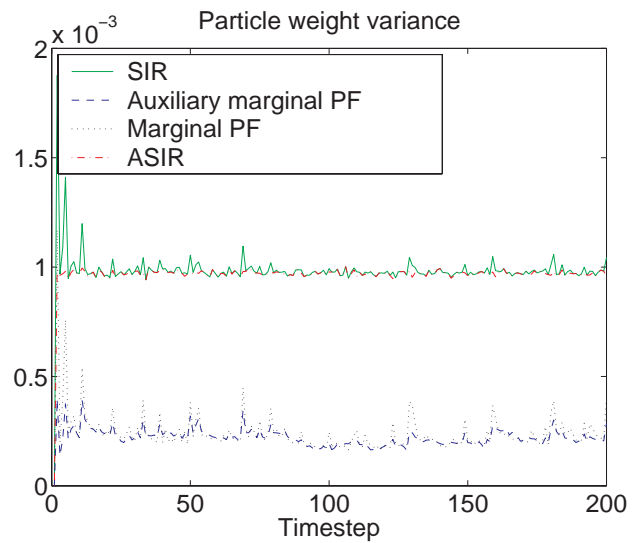

Figure 12: Stochastic volatility model; importance weight variance. Marginal filtering consistently achieves lower variance. See Figure 6 for a closer view of the first twenty timesteps, which more clearly illustrates the interaction among SIR, AMPF, and ASIR.

The results are summarized in Figures 11, 12, and 13. All results are means over five runs.

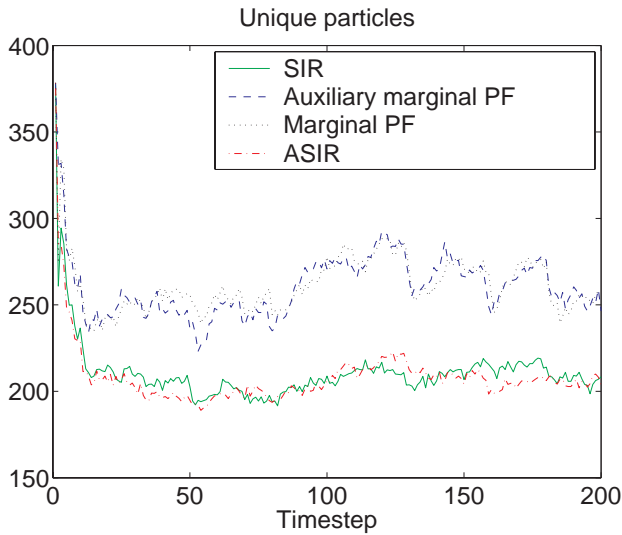

Figure 13: Stochastic volatility, unique particle count.

### 5.3 Fast implementation

We compared the MPF implemented naïvely to an implementation using the fast Gauss Transform (FGT), as it is conceivable that the error introduced by the FGT would offset the variance-reduction benefits that MPF provides. The results in Table 2 indicate that we can increase the precision sufficiently to render this issue moot. (Additionally, we performed all the experiments in the previous section using the approximation techniques.)

| $\epsilon$ | $N$ | | time(s) | speedup | RMSE |
|---|---|---|---|---|---|
| 1e-3 | 500 | naïve | 0.300 | | 1.2684 |
| | | FGT | 0.181 | **1.66** | 1.2685 |
| 1e-3 | 1500 | naïve | 2.568 | | 1.2469 |
| | | FGT | 0.310 | **8.28** | 1.2542 |
| 1e-7 | 5000 | naïve | 28.14 | | 1.2466 |
| | | FGT | 1.482 | **19.0** | 1.2466 |

Table 2: Fast methods applied to the marginal particle filter. All reported results are the mean values over ten runs. The FGT enables a substantial improvement in speed at the cost of a slight increase in error. For the last test, we substantially decreased the error tolerance of the FGT approximation (which increases the runtime), and were still able to achieve a considerable speedup and RMSE within the variance of the test to the naïve. This suggests that the error introduced by the FGT approximation can be made less than the inherent error caused by Monte Carlo variance.

## 6  Conclusions

Particle filtering involves importance sampling in the high-dimensional joint distribution even when the lower-dimensional marginal distribution is desired. We have introduced *marginal importance sampling* which overcomes this deficiency, and have derived two new particle filtering algorithms using marginal importance sampling that improve over SIR and ASIR, respectively. We have presented theoretical and empirical results which show that MPF and AMPF achieve a significant reduction in importance weight variance over the standard algorithms, and have shown how the computational burden can be drastically reduced.

Note that while we have reduced MPF to the same asymptotic complexity as SIR, the constants (while not prohibitive) are certainly higher than in SIR. The use of MPF is preferable when the transition prior $p(\mathbf{x}_t|\mathbf{x}_{t-1})$ is vague, which is common in industrial applications [10]. Regardless of the choice of prior, the MPF is also essential in order to compute the filter derivatives for parameter estimation [12].

**Acknowledgements**

We thank Gareth Peters for his invaluable suggestions.

## References

[1] C Andrieu, M Davy, and A Doucet. Improved auxiliary particle filtering: Application to time-varying spectral analysis. In *IEEE SCP 2001*, Signapore, August 2001.
[2] A Doucet, N de Freitas, and N J Gordon, editors. *Sequential Monte Carlo Methods in Practice*. Springer-Verlag, 2001.
[3] P Fearnhead. *Sequential Monte Carlo Methods in Filter Theory*. PhD thesis, Department of Statistics, Oxford University, England, 1998.
[4] A Gray and A Moore. 'N-Body' problems in statistical learning. In *NIPS*, pages 521–527, 2000.
[5] A Gray and A Moore. Nonparametric density estimation: Toward computational tractability. In *SIAM International Conference on Data Mining*, 2003.
[6] L Greengard and X Sun. A new version of the Fast gauss transform. *Documenta Mathematica*, ICM(3):575–584, 1998.
[7] S Kim, N Shephard, and S Chib. Stochastic volatility: Likelihood inference and comparison with ARCH models. *Review of Economic Studies*, 65(3):361–93, 1998.
[8] G Kitagawa. Monte Carlo filter and smoother for non-Gaussian nonlinear state space models. *Journal of Computational and Graphical Statistics*, 5:1–25, 1996.
[9] D Lang, M Klaas, and N de Freitas. Empirical testing of fast kernel density estimation algorithms. Technical Report TR-2005-03, Dept of Computer Science, UBC, February 2005.
[10] R Morales-Menendez, N de Freitas, and D Poole. Real-time monitoring of complex industrial processes with particle filters. In *Advances in Neural Information Processing Systems*, 2003.
[11] M K Pitt and N Shephard. Filtering via simulation: Auxiliary particle filters. *JASA*, 94(446):590–599, 1999.
[12] G Poyadjis, A Doucet, and S S Singh. Particle methods for optimal filter derivative: Application to parameter estimation. In *ICASSP*, 2005.
[13] C P Robert and G Casella. *Monte Carlo Statistical Methods*. Springer-Verlag, New York, 1999.
[14] C Yang, R Duraiswami, N A Gumerov, and L S Davis. Improved fast Gauss transform and efficient kernel density estimation. In *ICCV*, Nice, 2003.